\documentstyle[ApJ,times,psfig]{article}

\def\spose#1{\hbox to 0pt{#1\hss}}
\def\lta{\mathrel{\spose{\lower 3pt\hbox{$\mathchar"218$}}
     \raise 2.0pt\hbox{$\mathchar"13C$}}}
\def\gta{\mathrel{\spose{\lower 3pt\hbox{$\mathchar"218$}}
     \raise 2.0pt\hbox{$\mathchar"13E$}}}
\def\ergcm2s{{\rm erg\,cm^{-2} s^{-1}}}

\begin{document}

\title{Afterglow Emission from Naked Gamma-Ray Bursts}
\author{Pawan Kumar}
\affil{IAS, Princeton, NJ 08540}
\author{Alin Panaitescu}
\affil{Dept. of Astrophysical Sciences, Princeton University, NJ 08544}
\authoremail{pk@ias.edu, adp@astro.princeton.edu}

\begin{abstract}

We calculate the {\it afterglow} emission for Gamma-Ray Bursts (GRBs)
going off in an extremely low density medium, referred to as 
{\it naked bursts}. Our results also apply to the case where 
the external medium density falls off sharply at some distance from 
the burst. The observed afterglow flux in this case originates at
high latitudes, i.e. where the angle between the fluid velocity and
the observer line of sight is greater than $\Gamma^{-1}$. 
The observed peak frequency of the spectrum for naked bursts decreases 
with observer time as $t^{-1}$, and the flux at the peak of the 
spectrum falls off as $t^{-2}$. The 2--10 keV $X$-ray flux from 
a naked burst of average fluence should be observable by the 
SWIFT satellite for time duration of about $10^3$ longer than
the burst variability timescale. The high latitude emission contributes 
to the early $X$-ray afterglow flux for any GRB, not just naked bursts, 
and can be separated from the shocked inter-stellar medium (ISM) emission 
by their different spectral and temporal properties. 
Measurements of the high latitude emission could be used to map the angular 
structure of GRB producing shells.

\end{abstract}

\keywords{gamma rays: bursts -- gamma-rays: theory}

\section{Introduction}

A majority of long duration GRBs (lasting 10s or more) detected by 
the Dutch-Italian satellite BeppoSAX have detectable $X$-ray afterglows. 
The afterglow properties of shorter duration bursts is unknown, and it 
is possible that these bursts go off far away from galactic centers, 
where the ISM density is low. Will such bursts produce observable afterglows ?

The purpose of this paper is to show that all bursts, irrespective
of the ISM density, should have a detectable afterglow emission.
During the GRB the radiation is received from a region of the fireball of 
angular size $\Gamma^{-1}$ along the line of sight to the center of explosion. 
Emission from higher latitudes, $\theta>\Gamma^{-1}$, is received over a 
time interval that is long compared to the duration of the burst and, 
although this radiation is relativistically 
beamed away from the observer, it nevertheless has significant magnitude.
We calculate this emission and apply it to GRBs going off in a very low 
density ISM (\S 2). We also consider a case where the density of the 
circum-burst medium drops off abruptly at some radius (\S 3).

\section{High latitude emission from a relativistic shell}

Consider a spherical shell moving with Lorentz factor $\Gamma$.
The shell is shock-heated at some initial time and starts to radiate.
The emissivity in fluid rest frame, $\epsilon'_{\nu'}$, is a function of 
shell radius $r$ and we assume that it is independent of the angle 
$\theta$ relative to the observer's line of sight toward the center of 
the shell. We also assume that due to the radiative electron cooling and 
the adiabatic expansion of the shell (or some other process)
the emissivity in the observed energy band drops to zero when 
the shell radius is $r_c$. 
One can show in this case that the observed peak flux at observer 
time $t\gg r_c/\Gamma^2$ decreases as $t^{-2}$ and that the
observed peak frequency decreases as $t^{-1}$. 

A simple physical explanation for these results is the following. 
The flux per unit frequency from a relativistic source moving at 
an angle $\theta\gg\Gamma^{-1}$ is smaller by a factor of 
$(\theta\Gamma)^{6}$ compared to the case where the source 
is moving directly toward the observer 
(assume flat spectrum for simplicity). Integrating over sources 
located on equal arrival time surface we find the flux
ratio to be $(\theta\Gamma)^4$. The emission from angle $\theta$
arrives at a time which is larger than the photon arrival time from
a source at $\theta=0$ by a factor of $(\theta\Gamma)^{2}$.
Thus, the observer sees the flux falling off as $t^{-2}$. 
The observed frequency ratio in the two cases, for a fixed source 
frequency, is $(\theta\Gamma)^{-2} \propto t^{-1}$. 
A derivation for a more general case is given below.

The flux received at frequency $\nu$ from a shell moving
with Lorentz factor $\Gamma$ and velocity $v$ is given by
\begin{equation}
 f_\nu (t) = {1 \over 4\pi d^2} \int d^3r\, { \epsilon'_{\nu'}
                     ({\bf r}, t_{lab}) \over \Gamma^2 (1-v\mu)^2 } \;,
\label{fnu1}
\end{equation}
where $\mu=\cos\theta$, $\theta$ is the angle between fluid velocity and
the line of sight to the observer, and $\nu'=\nu\Gamma(1-v\mu)$ \& 
$\epsilon'_{\nu'}$ 
are the frequency and emissivity in the shell rest frame at radius $r$ and 
laboratory frame time $t_{lab}=t_+ r\mu$. For a shell of thickness $\Delta r$
(in the lab frame) much smaller than $r$, the angular integration in the
above equation can be carried out to yield 
\begin{equation}
 f_\nu = {1\over 2 d^2} \int dr \, r {\Delta r\epsilon'_{\nu'}(r)
          \over \Gamma^2(1-v\tilde{\mu})^2} \;,
\label{fnu2}
\end{equation}
where $\tilde{\mu}(r) = (t_{lab}-t)/r(t_{lab})$. 

Let us assume that the observed spectrum during the GRB phase, i.e. before 
the shell cools, is a power-law function of index $\beta$
between the observed energy band and the peak of the spectrum:
$\epsilon'_{\nu'} = \epsilon'\nu'^{\beta}$, and that, as $\theta$ is 
varied, the co-moving frame observing frequency does not pass through
any breaks. In this case we can rewrite equation (\ref{fnu2}) as
\begin{equation}
 f_\nu = {\nu^{\beta}\over 2d^2} \int dr\, {\Delta r\epsilon' r^{3-\beta}
   \over \Bigl[ \Gamma(t + r/v - t_{lab})\Bigr]^{2- \beta} } \;.
\label{fnu3}
\end{equation}

As mentioned above, we consider the case where the injection of accelerated 
electrons stops at a certain radius $r_c$. This can happen 
either because the internal shock has finished traversing the shell or 
because the density of the ISM drops by a large factor. 
In both cases the electrons undergo adiabatic cooling beyond $r_c$ which
leads to a sharp fall off of $\epsilon'$ (see \S 3); for radiative electrons
$\epsilon'$ drops off even faster.

The integrand in equation (\ref{fnu3}) is a rapidly increasing function 
of $r$, hence most of the contribution comes from $r\sim r_c$, 
and the peak flux is given by
\begin{equation}
 f_{\nu_p}(t) \approx f_{\nu_p}(t_c) \left( {2\tau \over 
         t +\tau - t_c } \right)^2 \;\;, 
         \tau \equiv {r_c \over 2 \Gamma_c^2} \;,
\label{fnum}
\end{equation}
where $t_c = t(r_c)$ and $\Gamma_c \equiv \Gamma (r_c)$.
The peak frequency $\nu_p$ of the observed flux decreases with time as
\begin{equation}
 \nu_p (t) = \nu_p(t_c) \left( { \tau \over t +\tau - t_c } \right) \;.
\label{num}
\end{equation}

For a shell that is energized by the collision with another shell
(i.e. internal shocks) and expands in vacuum, $\tau$ is also a measure 
of the duration $\delta t$ of the pulse emitted by the shell. 
For $t \gg t_c$, equations (\ref{fnum}) and (\ref{num})
become
\begin{equation}
  f_{\nu_p}(t) = f_{\nu_{p,c}} \left( { \delta t \over t } \right)^2 \;\;, 
 \nu_p(t) = \nu_{p,c} \left({\delta t \over t}\right)\;.
\label{fnum1}          
\end{equation}
where $f_{\nu_{p,c}} \equiv f_{\nu_p}(t_c)$ and $\nu_{p,c} \equiv \nu_p(t_c)$. 
For power-law spectra, the observed flux at a frequency $\nu$ can be calculated
from equations (\ref{fnu3}) \& (\ref{fnum1}): 
\begin{equation}
 f_\nu (t) = f_{\nu_{p,c}} \left( { \delta t \over t } 
 \right)^{2-\beta} \left( { \nu \over \nu_{p,c} } \right)^\beta \;.
\label{fnu4}
\end{equation}

For a GRB consisting of $N$ pulses the high latitude afterglow flux is 
the sum of flux from each pulse given in equation (\ref{fnu4}). 
For an average peak amplitude of $\overline{f}_{\nu_p}$, the afterglow flux
is approximately equal to $\overline{f}_{\nu_p} N^{\beta-1}
(t_G/t)^{2-\beta}$; where $t_G\sim N \delta t$ is the duration of the GRB.

The low energy power-law index for GRBs is in the range of $-1$ and 2 
with the peak of the distribution at $\sim 0$, and the high energy 
index is between $-3$ and $-0.3$ with peak at $-1.2$ (Preece et al. 2000).
Thus the low energy light-curve for naked bursts
is expected to fall-off as $t^{-2}$, whereas the light-curve
at high energy should decline as $t^{-3.2}$. For synchrotron emission
the spectral index below the peak is $\beta=1/3$ and the afterglow from
high latitude emission is expected to decline like $t^{-5/3}$
until the synchrotron peak passes through the observing band.

As an example, consider a naked GRB lasting for 10 seconds and with a mean flux 
of $10^{-6} \ergcm2s$ and the spectral peak at a few hundred keV. At 
$10^3$ seconds after the burst the spectral peak is in the 2--10 keV band 
and the flux is $\sim 10^{-12} \ergcm2s$ (see Figure 1). 
Assuming that the optical emission is not self-absorbed, the optical 
flux at $t = 300$ s corresponds to $R \sim 25$. 

The high latitude emission should be detectable by the SWIFT satellite 
as an $X$-ray afterglow following short duration GRBs, which are perhaps
produced as a result of neutron star merger in a low density medium. 
Some of the early $X$-ray afterglows observed by BeppoSAX have a power-law 
decay of index of $\gta 1.6$, which could have had a contribution from
the high latitude emission. In the few cases where an $X$-ray afterglow 
was continuously monitored for $10^3-10^4$ seconds after the main burst
(i.e. GRB 910402 -- Tkachenko et al. 2000, GRB 920723 -- Burenin et al. 
1999, and GRB 980923 -- Giblin et al. 1999), the $X$-ray light-curve 
exhibited a decay significantly slower than what is expected from the 
high-latitude emission, implying that the emission from the external 
shock must have been dominant from very early times.

The ratio of the observed flux from the shocked ISM gas and the high 
latitude emission ($Q$) depends on burst parameters and, 
generally, on the density of the ISM. An important time time scale for
this comparison is the shell deacceleration time $t_{da}\approx 
100 E_{52}^{1/3} (1-\eta)^{1/3} (\eta n_0\Gamma_2^8)^{-1/3}$ sec; 
where $E_{52}$ is isotropic equivalent of energy in observed 
gamma-ray emission in units of $10^{52}$ erg, $\Gamma_2 = \Gamma_0/100$, 
$\Gamma_0$ is the initial Lorentz factor of ejecta, and
$\eta$ is the efficiency factor for converting energy in explosion
to the observed gamma-ray emission. We consider $t_{da}$ less than 
or greater than the GRB duration, $t_G$, separately below.

Let us first consider $t_{da}\lta t_G$.
For a ISM density ($n_0$) larger than $10^{-2}\; {\rm cm^{-3}}$ and for 
$\epsilon_B \sim 10^{-2}$, the soft $X$-ray domain is above the cooling 
frequency after the GRB, and in this case $Q$ is independent of $n_0$. For 
an electron energy index $p=2$, $Q \sim 0.3 (\nu_G/\nu)^{\beta+1}
\epsilon_e (t/\delta t_G)^{1-\beta}(1-\eta) \eta^{-1}$, where 
$\nu_G$ is the observed peak of the GRB spectrum, $\nu$ is the frequency 
for the afterglow observation, and $\epsilon_e$ is the energy fraction 
in electrons in external shock. For $\nu_G/\nu = 20$, $\epsilon_e = 
0.1$, $\beta = 0$, and $\eta=0.1$, the emission from the external shock 
is larger by about a factor of 5 at the end of the GRB. This result has 
a weak dependence on $p$. 

For $t_{da}\gg t_G$, as is expected for low density
ISM, and the observing frequency smaller than the cooling frequency,
and for $p=2$ \& $\beta=1$, we find $Q\approx 8 \epsilon_B^{3/2}
\epsilon_e E_{52}^{1/4}n_0^{1/2} t_{da}^{1/4} (\nu_G/\nu)^{2}\nu_{10}^{1/2} 
(1-\eta)^{5/4}\eta^{-5/4}$;  $\nu_{10}$ is frequency in units
of 10 keV. As an example, for $\epsilon_e = 0.1$, $\epsilon_B = 10^{-2}$,
$\eta = 0.1$,  $\nu_G/\nu = 20$, $\nu_{10}=1$, $E_{52}=1$ and 
$t_{da}=100$s, the two emissions are equal 
at the deacceleration time for $n \sim 4\times10^{-3}\; {\rm cm^{-3}}$. 
For $t_{da}\gg t_G$, $\nu$ greater than the cooling frequency,
and $\beta=1$, 
$Q=2\epsilon_e(t/t_{da})^3 (\nu_G/\nu)^{2} (1-\eta)\eta^{-1}$.

One should be able to separate out the contributions of the high latitude 
and shocked gas emissions by using the difference in their spectra and 
light-curve slopes: the $X$-ray spectra for shocked low density ISM 
is $f_\nu \propto t^{-3(p-1)/4} \nu^{-(p-1)/2}$ whereas the 
high-$\theta$ spectrum is the low energy part of the GRB spectrum, i.e. 
$f_\nu \propto t^{-(2-\beta)} \nu^{\beta}$ with $-1 <\beta < 2$.

\section{Fireball expansion into ISM with a density discontinuity}

In this section we consider an external shock propagating in a ISM which 
consists of two regions of different densities.
The model we consider consists of a fireball that
shock the interstellar medium, producing a standard afterglow
emission, and then at some radius $r_{ad}$ the 
density of the medium drops precipitously and the shell  
expands adiabatically so that
the thermal energy of protons, electrons and magnetic field
is converted back to the bulk kinetic energy of the shell.
We follow the shell evolution and synchrotron radiation
starting from the time of the free expansion of the shell.

The thermal Lorentz factor of particles, in an adiabatically expanding shell, 
decreases as $\gamma_{th}\propto V^{-1/3}$, and the bulk Lorentz factor of 
the shell ($\Gamma$) increases with time as $V^{1/3}$; where 
$V=\pi\theta_0^2 r^2\Delta r$ is the co-moving volume of the shell, 
and $\Delta r\propto r^{1/4}$ is the co-moving shell thickness.  

We consider the collimation angle $\theta_0$ of the ejecta to be constant
in which case the thermal Lorentz factor decreases with 
$r$ as $r^{-3/4}$, and the bulk Lorentz factor of the shell 
increases as $r^{3/4}$. Therefore
the evolution of the thermal Lorentz factor is given by
\begin{equation}
\gamma_{th} \approx \Gamma_{ad} \left( {9\over 8} -
   {t\over 8 t_{ad}} \right)^{3/2},
\label{gamth}
\end{equation}
where $\Gamma_{ad}$ and $t_{ad}$ are the bulk Lorentz factor and
the observer time respectively at the onset of the free adiabatic expansion.
The bulk Lorentz factor $\Gamma\approx \Gamma^2_{ad}/\gamma_{th}$.

The magnetic field strength, assuming that it is tangled, decreases
as $V^{-2/3}\propto\gamma_{th}^2$. Thus the peak synchrotron frequency,
in the observer frame, scales as $\gamma_{th}^3$, and the peak flux
$f_{\nu_p}\propto \gamma_{th}$. The flux at a frequency greater
than the synchrotron peak $\nu_m$ but smaller than the cooling frequency
$\nu_c$ is given by
\begin{equation}
f_\nu\propto \left( {9\over 8} - {t\over 8 t_{ad}}
    \right)^{3(3p-1)/4} \;,
\label{fnue}
\end{equation}
and the power-law index $\alpha=-d\ln f_\nu/d\ln t$ is 
\begin{equation}
 \alpha={3(3p-1)(t/t_{ad})\over 4(9-t/t_{ad})}\;.
\label{bet}
\end{equation}

Therefore the afterglow light-curve steepens continuously; in the beginning 
of the adiabatic expansion $\alpha=3(p-1)/4$ while at $t = 3 t_{ad}$,
$\alpha=3(3p-1)/8$, assuming that $\nu_m < \nu < \nu_c$. 
As the light-curve slope increases with time, the flux from 
higher latitudes takes over. For a shell interacting with a uniform 
circum-burst material at $r < r_{ad}$, $\Gamma \propto r^{-3/2}$, and thus 
$t_{ad} = r_{ad}/8\Gamma_{ad}^2$. For $t \gg t_{ad}$
equation (\ref{fnum}) gives
\begin{equation}
 f_\nu (t) = 16\, f_{\nu_{p,ad}} \left( { t_{ad} \over t } 
       \right)^{2-\beta} \left( \frac{\nu}{\nu_{p,ad}} \right)^\beta \;,
\end{equation}
therefore the high latitude emission prevents $\alpha$ from becoming 
larger than $(p+3)/2$.

These results apply over a limited range of $t$. 
At late times the non-zero density of the ISM prevents 
the free expansion of the shell and the freshly shock-heated 
gas contributes to the observed flux. The free expansion of the
shell is terminated when the mass of the swept-up low density gas
is $\sim E/\Gamma_{ad}^2$, $E$ being the energy of the adiabatic shell.
Thus the radius at which the free adiabatic expansion is terminated
is $r/r_{ad}\sim (n_1/n_2)^{2/9}$, where $n_1$ and $n_2$ are the
densities of the high and low density ISM respectively. The time in
the observer frame when the adiabatic expansion ends is 
\begin{equation}
 {t_f \over t_{ad}} \approx 9 - 8\left({n_2\over n_1}\right)^{1/9}.
\label{tob}
\end{equation}
For $n_2/n_1=$ 0.1 (0.01) free expansion is terminated at $t_f/t_{ad}=2.8$ (4.2).

The value of $\alpha$ reverts back to $3(p-1)/4$ when the emission 
from shocked low-density ISM takes over. At the time when the adiabatic
expansion of the shell ends the ratio of the flux from 
the low density shocked gas to the flux at $t_{ad}$ is
$\sim (n_2/n_1)^{(7-p)/12}$. If the fractional energies in electrons, 
$\epsilon_e$, and magnetic field, $\epsilon_B$, are same for shock in
the high and low density ISM then the flux at $t_f$ due to
high latitude emission and the low density shocked gas are approximately
equal for $10^{-3} < n_2/n_1< 10^{-1}$. Since
the observed flux for frequency between $\nu_m$ and $\nu_c$
is proportional to $\epsilon_e^{(p-1)} \epsilon_B^{(p+1)/4}$,
values of $\epsilon_e$ \& $\epsilon_B$ for the shocked lower density 
medium smaller by a factor of 10 could reduce the flux 
from the low density shock gas so that the high-latitude emission 
dominates for $\sim 20\, t_{ad}$, and during this period 
the power-law index of the light-curve is $\alpha = (p+3)/2$; for 
$\nu>\nu_c$, $\alpha = (p+4)/2$.

The optical light-curve of the afterglow of GRB 000301C fell off as 
$\sim t^{-1}$ for the first three days and subsequently steepened
to $\sim t^{-3}$ (Rhoads \& Fruchter 2000). 
 From simultaneous optical--IR observations Rhoads \& Fruchter (2000) and
Sagar et al. (2000) have found that $\beta =-0.9\pm0.1$ at $t \sim 4$ days.
A possible explanation for the steep decay seen at late time in this afterglow 
is that the ISM density fell off at some radius\footnote{
  The optical emission of the afterglow of GRB 000301C shows considerable 
  variability prior to the steepening of the light-curve decline. 
  This suggests that there are significant fluctuations in the ISM density,
  and so it is not altogether surprising that the external medium density drops 
  to a small value at some radius.} 
and the subsequently
observed afterglow emission arose at $\theta \gg \Gamma^{-1}$, yielding a 
power-law decaying light-curve of index $\alpha=2-\beta=2.9\pm0.1$,
which is consistent with the data.

Figure 2 shows the observed $R$-band data for GRB 000301C and the 
theoretically calculated light-curve based on the model described here.
The transition time for light-curve steepening is $\sim 10\; t_{ad}$, 
which is roughly consistent with the observations. 
It has been shown by Kumar \& Panaitescu (2000) that the time-scale for 
light-curve steepening due to jet edge effects in a homogeneous ISM is 
roughly comparable. The late time power-law index according to the jet model 
is $\alpha = 1-2\beta=2.8\pm0.2$ for adiabatic electrons radiating at optical 
frequencies, which is also consistent with the data. The different relationship
between $\alpha$ and $\beta$ in these two models can be used to distinguish
between them.

\section{Conclusions}

The main conclusion of this work is that gamma-ray bursts going off
in vacuum -- {\it naked} GRBs -- should have $X$-ray afterglow emission 
detectable by the $X$-ray telescope aboard SWIFT satellite for about an hour
after the GRB. 
This radiation originates at the high latitude, $\theta\gg\Gamma^{-1}$,
part of the gamma-ray emission surface. The flux in a fixed observer 
energy band below the peak falls off as $t^{-5/3}$, while the peak 
flux decreases as $t^{-2}$. The peak frequency of the observed flux 
falls-off as $t^{-1}$.

For a burst going off in a non-zero density ISM the early afterglow
flux, within the first hour, is the sum of emission from high latitude 
and the shocked ISM. The two can be distinguished based on the differences 
between their spectral and temporal slopes; the $X$-ray spectrum for the 
shocked ISM is $\sim \nu^{-1}$, whereas for the high latitude radiation the
spectrum should be the same as the GRB spectrum at low energies, i.e. 
$\sim\nu^{0.3}$. The measurement of the high latitude afterglow emission
should help map the irregularities in the ejecta producing the GRB and their 
collimation before these are detected in the emission from the shocked ISM.

The radiation emitted from latitudes $\theta \gta \Gamma^{-1}$ sets an 
upper bound on the steepness of the flux decline: we expect the 
observed $\gamma$-ray flux for each individual peak within
the burst to fall off less rapidly than $t^{-(2-\beta)}$, 
where $t$ is measured from the peak of the pulse
and $\beta$ is the spectral index ($f_\nu \propto \nu^\beta$). 
A more rapid flux decline would be an indication of either an extremely
small jet opening-angle or a very inhomogeneous shell, as in the model 
suggested by Kumar \& Piran (2000).

Another straightforward consequence of the high latitude emission 
is that the power-law decline for the afterglow light-curve
can not be larger than about 3, even when the fireball expands into vacuum.
The observed late time power-law index for the light-curve of GRB 000301C 
is about 3, which is larger by about 2 compared to the early time index.
This large and rapid steepening of the light-curve could arise when
the late time light-curve is dominated by emission from high latitudes.

\acknowledgments{We thank Bohdan Paczy\'nski and Tsvi Piran for useful 
discussions.}

\clearpage

\begin{figure*}
\vspace*{-2cm}
\centerline{\psfig{figure=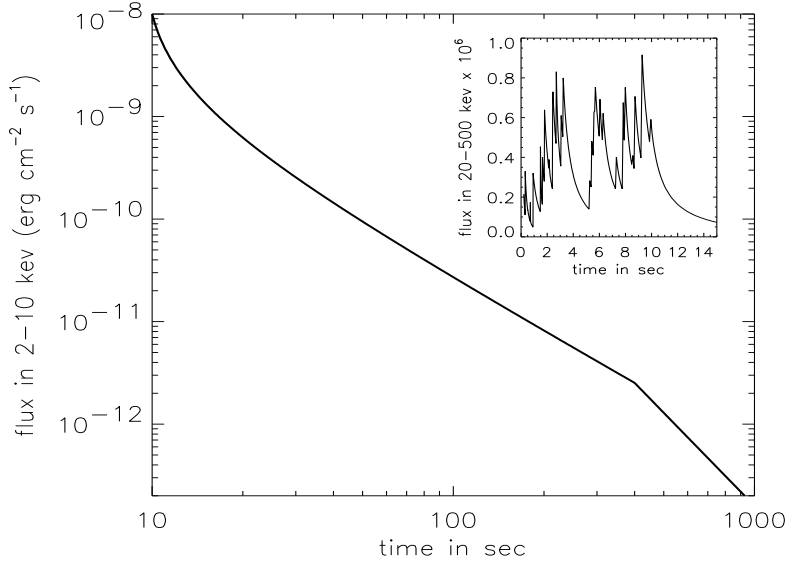,height=8cm,width=10cm}}
\figcaption{2--10 keV light-curve arising from high latitudes, i.e. 
$\theta>\Gamma^{-1}$. The ISM density is assumed zero. The peak 
flux for the GRB is taken to be $10^{-12} \ergcm2s {\rm eV^{-1}}$; 
the burst duration is 10 seconds, the low energy slope of the spectrum
($\beta$) is 1/3 and the high energy index is $-1$. The burst duration and the 
peak flux set the x- and y- axis scales, and the spectral slopes set 
the power-law decline of the light-curve (see text). The power-law index 
of the light-curve is 5/3 for $t<400$s and steepens to 3 when the 
peak frequency passes through the x-ray band. The inset shows the 
GRB light curve in 20--500 keV band, the mean pulse width is 0.4s.} 
\end{figure*}

\begin{figure*}
\vspace*{-2cm}
\centerline{\psfig{figure=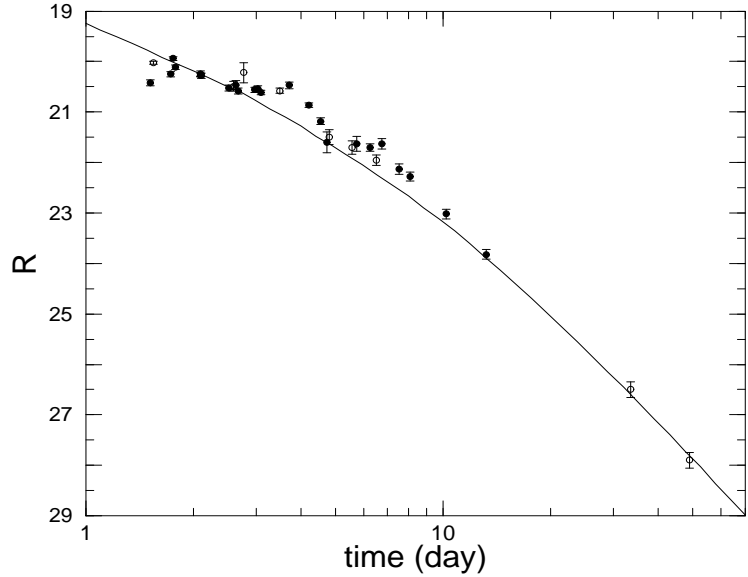,height=8cm,width=10cm}}
\figcaption{$R$-band light-curve (continuous line) from a spherical remnant 
running into a uniform density ISM which ends at some radius. Beyond this 
radius, at 3 day in the observer frame, the shell undergoes adiabatic 
expansion. Before the break the optical flux decay slope is $3(p-1)/4$, 
and after the break it is $(p+4)/2$ for frequencies above the cooling break.
The parameters for the model are: $E=2\times 10^{52}$erg, $p=2.6$, 
$n=1$ cm$^{-3}$, $\epsilon_B=5\times10^{-3}$, $\epsilon_e=0.1$. 
The cooling frequency crosses $R$-band around 5 days.
The data is taken from Sagar et al. (2000), Massetti et al. (2000)
and from GCN Circulars: Bernabei et al. 2000, Bhargavi \& Cowsik 2000, 
Castro-Tirado et al. 2000, Fruchter et al. 2000, Fynbo et al. 2000, 
Gal-Yam et al. 2000, Garnavich et al. 2000, Halpern et al. 2000, 
Mujica et al. 2000, Veillet et al. 2000 .}
\end{figure*}

\end{document}